# Characteristics of multiplexing parallel beam (MPB) by using Monte Carlo simulation

ZHAO Jing-Wu(赵经武) MA Jia-Yi(马加一) SHI Xiao-Dong(史晓东)

School of Physics, Nanjing University, Nanjing 210093, China

**Abstract:** It is discussed that the point spread function (PSF) of multiplexing parallel beam (MPB) is shift-invariant; the spatial resolution is linear as a function of depth $h$ and the geometrical sensitivity is converged to a constant value at far field, by using Monte Carlo simulation in this work. The results indicate that the MPB collimator may maintain the same system characteristics as parallel beam (PB) collimator. Although the minimum standard deviation (SD) of MPB is more 3.7% than the PB with high sensitivity (HSPB), the improved resolution is 55.2% at $h=160mm$, compared with HSPB in this work. The reconstruction process also indicates that the convergence rate of MPB is slower than the HSPB because the projection counts are part of overlapping and the PSF of oblique PB (OPB) is asymmetrical. However the effective hole length is used in the calculation of the system matrix to correct for the penetration, which has a satisfactory result. From a visual assessment, the reconstructed image may well satisfy clinical criteria and it is therefore worth investigating this novel MPB collimator system for clinical imaging.

## 1. Introduction

Molecular imaging with single photon emission computed tomography (SPECT) is a suitable tool to clinically study abdominal and thoracic diseases such as tumors, infection and bone diseases. Although image quality of SPECT is not better than PET, in China it is a quite useful technique because it is cheaper than PET. In SPECT, parallel-beam (PB) collimator with a large field of view (FOV) is known as one of main collimators. However, the PB collimator suffers from low spatial resolution when SPECT with high sensitivity is used to clinical imaging. On the contrary, it suffers from low sensitivity when the resolution is high. Therefore there is tradeoff between the sensitivity and resolution. The tradeoff limits the PB collimator application in clinical imaging. At present the SPECT only plays a complementary role in medical imaging, because of low resolution or high statistical noise. It not precisely locates disease sites. To improve the property of PB collimator, some methods were proposed. For example, the advantage of pixilated semiconductor detectors [1]-[3] and hybrid collimator was applied [4]. The author [5] also indicates

E-mail: jwzhao@nju.edu.cn

the lack of hybrid collimator. While simultaneously using a parallel-beam collimator and a cone-beam collimator, the parallel-beam collimator can provide sufficient sampling and the cone-beam collimator is able to provide higher photon counts. But this hybrid system does not give significant improvement in terms of image noise and artifacts reduction.

We proposed a multiplexing PB (MPB) collimator in previous work [6] [7]. By introducing a multiplexing element, which is introduced in each transverse slice, into the PB collimator, the work had shown that the MPB is in their ability to improve the tradeoff between the sensitivity and resolution. Because of resulting in a set of gamma ray channels (GRCs) in the transverse direction, photons from three different directions can also reach the detector unit (bin). Therefore, the MPB collimator is able to provide more photon counts. The performance of MPB is just like as consisting of three parallel beams. Based on the previous work, some characteristics of the MPB is investigated and compared with the PB in this paper, by suing Monte Carlo simulation.

**2 Method**

2.1 Sensitivity compensation

In SPECT imaging, when the hole length of PB collimator with high sensitivity (HSPB) is increased in order to improving resolution, the sensitivity is decreased. As shown in Fig.1, LS is loss of sensitivity. The dash line shows the increased hole length. When the increased septa are cut off and pulled up to make GRCs, the LS can be compensated by oblique PB ($OPB_1$ and $OPB_2$). The GRCs is introduced in the centers of the septa in each transverse slice. While making appropriately the GRCs width $w$, the sensitivity of MPB collimator can maintain the same sensitivity of HSPB. And the resolution of MPB collimator is higher than the resolution of HSPB because of increased hole length. The line 1 is condition of maintaining three PB. The $d$ is hole diameter. The $t$ is septa thickness. The $l_e$ is hole length of the HSPB. The $l$ is hole length of the MPB. The $\theta$ is oblique angle of OPB. The W is area of received photons when the GRC is considered as oblique PB.

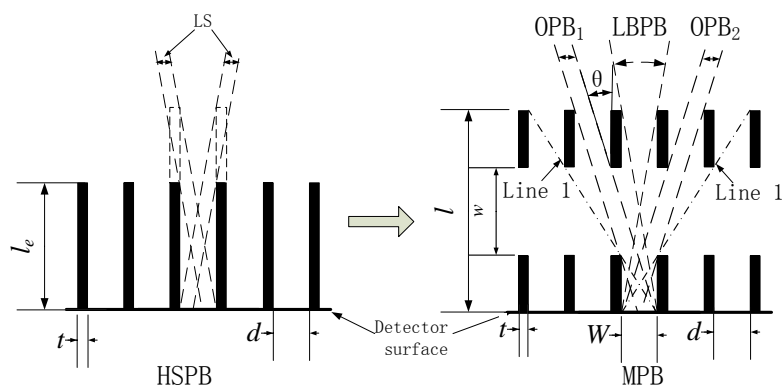

Fig.1 Schematic diagram of sensitivity compensation

## 2.2 Monte Carlo simulation

In this work, all the simulation environments are modeled using Monte Carlo N-Particle Code (MCNP4C) [8]-[10]. In the Monte Carlo simulation, the septa material is tungsten. The scintillator is placed on the back surface of the collimators. The detector effect is set to 1. The data were recorded using a 10% energy window centered on 140 keV (126–154 keV). All of the photon tracks that reached the detector were recorded and analyzed.

To simulate the point spread function (PSF), point sources with depths $h$ (i.e., the distances from the point source to the face of the collimators) from 1 cm to 32 cm with 1 cm increments are placed in the center of the collimators, just over the central bin. The energy of gamma rays is 140 keV ($^{99m}$Tc), and 2 billion photons were emitted over the $4\pi$ solid angle.

Based on the PSF, the geometrical sensitivity and spatial resolution were evaluated. The geometrical sensitivity of collimators is the sum of the geometrical effect of all detector bins. The spatial resolution was evaluated by the full width at half maximum (FWHM) of the PSF in previous work [6]. However, the FWHM in the case of MPB may mislead due to the disjointed nature of its PSF. Thus another metric, the root mean squared (RMS) resolution [11], may be an appropriate way to evaluate MPB in this work. RMS is given as follows:

$$RMS = \sqrt{\overline{v^2} - \overline{v}^2} \quad (1)$$

where $v$ is the bin position. The average value $\overline{v}$ and $\overline{v^2}$ of $v$ are weighted by the PSF. Because the PSF of MPB is constituted of PB, OPB$_1$ and OPB$_2$, the RMS of MPB can be estimated as:

$$RMS_{MPB} = \sqrt{RMS_{PB}^2 + RMS_{OPB_1}^2 + RMS_{OPB_2}^2} \quad (2)$$

The RMS of PB, OPB$_1$ and OPB$_2$ can be respectively calculated by equation (1).

As there is no GRC effect in the axial direction [6], to simplify the simulations, we simulate 2D image reconstruction using a 2D mathematical phantom, as shown in Fig. 2. The phantom is placed perpendicularly to the axis of rotation (AOR) with the AOR through the center of the phantom. There are four uniform cylindrical hot spots with background with diameters of 3.0, 2.0, 1.0 and 0.25 cm. The diameter of the background is 30 cm. The spot-to-background ratio is 5:1. The energy of gamma rays is 140 keV ($^{99m}$Tc), and 1.06 billion photons were emitted over the $4\pi$ solid angle for each projection. In the phantom simulation, 120 projections from 0° to 357° with 3° steps were acquired. The parameters of the MPB and HSPB for Monte Carlo simulation are given in Table 1.

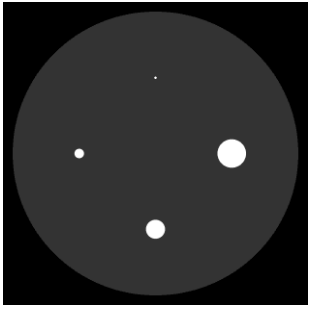
Fig.2 True phantom for Monte Carlo simulation

Table 1 The geometric parameters of the MPB collimators

| $d$ (cm) | $l$ (cm) | $l_e$ (cm) | $ROR$ (cm) | $D^t$ (cm) | $D^a$ (cm) |
|---|---|---|---|---|---|
| 0.209 | 4.9 | 3.57 | 16.0 | 38.0 | 38.0 |
| Bins$^t$ | Bins$^a$ | $w$ (cm) | $\theta$ (°) | $t$ (cm) | |
| 151 | 151 | 1.8 | 4.29 | 0.0424 | |

$d$: Hole diameter; $l$: Hole length of MPB; $l_e$: Hole length of HSPB; $ROR$: Radius of rotation; $D^t$: Total transverse length; $D^a$: Total axial length; Bins$^t$: Bins (transverse direction); Bins$^a$: Bins (axial direction); $w$: Width of GRC; $\theta$: Oblique angle $\theta$; $t$: Septal thickness.

**3. Result and Discussion**

3.1 Shift-invariant

Images of the PSFs of the HSPB and MPB obtained by simulation at AOR (160 mm) are shown in Fig. 3. The PSF of the HSPB has a distribution in a cross shape with a single peak at the centre. The PSF of the MPB also has a distribution in a cross shape. However, the crosswise is longer than the vertical. And there are three peaks on crosswise. The three peaks are respectively contributed by a PB and two OPB [6]. Profiles of the PSFs obtained by simulation are shown in Fig.4. The markers show the results from Monte Carlo simulation. The circle is the PSF of MPB and the star is the HSPB.

For PB collimator, the view spread function (VSF) image coincides with the PSF images of the PB [1]. Therefore, the PSF of PB is shift-invariant. The blurring of the VSF is essentially the same as that of the PSF. By contrast, the PSF of MPB is linear superposition of PB, OPB$_1$ and OPB$_2$. And the PSF of OBP is asymmetrical and as similar as the SPF of slant hole collimator [12]. So the PSF of OBP is also shift-invariant. Therefore the PSF of MPB is also shift-invariant. The VSF image of MPB coincides with the PFS of MPB, although there is difference in the concepts of the PSF, compared with the HSPB.

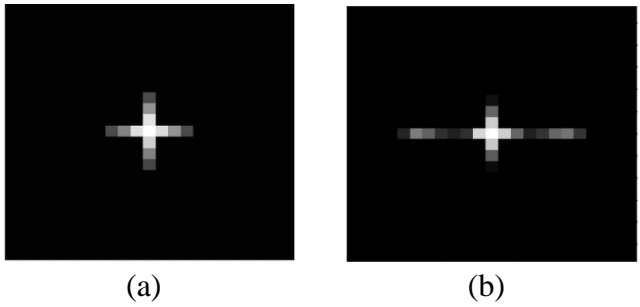

(a)             (b)

Fig.3 The PSFs Image of (a) HSPB and (b) MPB

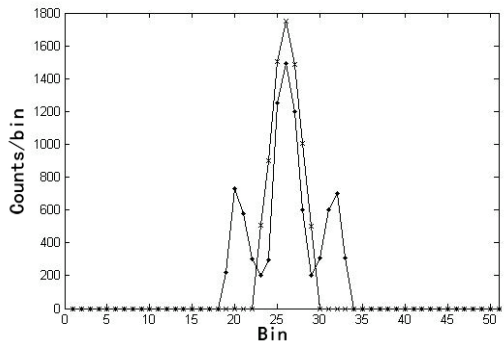

Fig.4 The profiles of the PSFs of HSPB and MPB

3.2 Spatial resolution

As shown in Fig.4, the PSF of MPB is linear superposition of PB, $OPB_1$ and $OPB_2$. And the PSF of MPB is shift-invariant, so that the blurring of the VSF is essentially the same as that of the PSF. The resolution is contributed by the PSFs of PB, $OPB_1$ and $OPB_2$. And the PSF of OPB is asymmetrical. Therefore it is appropriate that the resolution is evaluated by the RMS. Fig. 5 shows the RMSs of MPB and HSPB as a function of depth $h$ in the transverse direction.

The RMSs of HSPB and MPB are all linear as a function of depth $h$ because the two collimators are shift-invariant. But the MPB is obviously better than the HSPB. At AOR (160 mm), the improved resolution is about 55.2% in this wok.

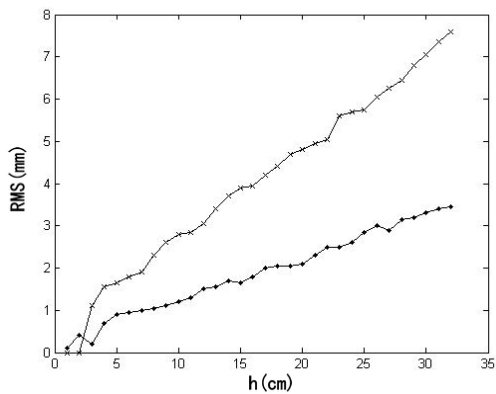

Fig.5 The RMSs of HSPB and MPB as a function of depth $h$.

3.3 Geometrical sensitivity

Based on the simulation results, the geometrical sensitivity is calculated as the sum of the geometrical effect for each detector bin. Fig.6 shows the geometrical sensitivity as a function of depth $h$. The MPB is constituted of PB, $OPB_1$ and $OPB_2$. The geometrical sensitivities of HSPB and MPB should be same variation of shape as depth $h$. The oscillation of sensitivity is because hole numbers of received photon is increased as increasing depth $h$. And the sensitivity is converged to a constant value at far field, which is an obvious character for parallel beam in SPECT image. The constant values of HSPB and MPB is almost equal to each other in this work.

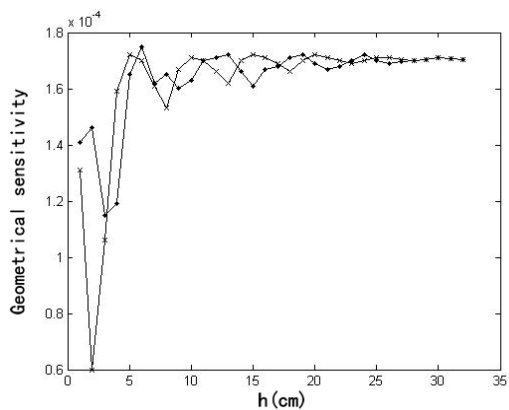

Fig.6 The geometrical sensitivity as a function of depth $h$.

3.4 Image reconstruction

In the total counts (about $1.9 \times 10^5$ per projection angle) acquired with MPB collimator for the Monte Carlo simulation of the phantom shown in Fig. 2, the penetration is about 3.7%, and the scattering is about 0.71%. In the total counts (about $1.7 \times 10^5$ per projection angle) acquired with HSPB collimator for the Monte Carlo simulation of the phantom shown in Fig. 3, the penetration is about 1.8%, and the scattering is about 0.93%.

Ordered subset expectation maximization (OSEM) reconstruction, which takes into account the Poisson noise of the measured data, is used for 2D image reconstruction, and the system matrix is obtained through ray tracing [13]. The penetration is corrected by using effective hole length and width of GRC [14]. For HSPB, $l_{e,eff} = l_e - 2/\mu$. For MPB, $l_{eff} = l - 2/\mu$ and $w_{eff} = w - 2/\mu$.

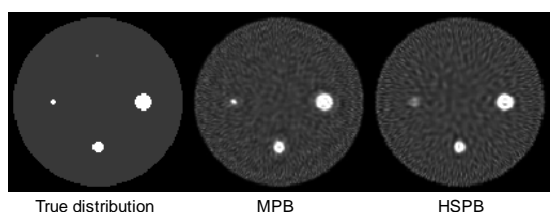

True distribution     MPB     HSPB

Fig.7 The reconstructed image

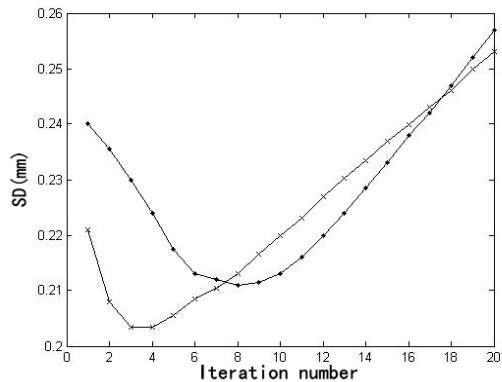

Fig.8 The SD as a function of iteration number

Fig. 7 shows the reconstructed image of MPB and HSPB collimators with 10 subsets at iteration numbers 15 and 14 respectively, which have same noise level. Each reconstructed image has a resolution of 128 ×128 pixels, with a pixel size of 2.5 mm. Fig. 8 shows the standard deviation (SD) as the iteration number.

The counts acquired with MPB indicate that the increased penetration is about 106% and the decreased scattering is 60.6%, being relative to HSPB collimators. The increased penetration is not 200% because GRC is at central of hole length. The decreased scattering is due to decreased area of septa by introducing GRC. The Fig.7 indicates that the resolution of MPB is better than the HSPB. But the Fig.8 indicates two problems: (1) the convergence rate of MPB is slower than the HSPB because the projection counts are part of overlapping and the PSF of OPB is asymmetrical. And (2) the minimum SD of MPB is more 3.7% than the HSPB. However, the cost of increased SD is worthy, compared with the improved resolution 55.2%. The convergence rate can be improved by increasing subset number.

## 4. Conclusion

Using Monte Carlo simulation in this work, we demonstrate that the PSF of MPB is shift-invariant, the resolution is linear as a function of depth $h$ and the sensitivity is converged to a constant value at far field. Therefore, the MPB collimator may maintain the same system characteristics as HSPB collimator. Although the minimum SD of MPB is more 3.7% than the HSPB, the improved resolution is 55.2%. The reconstruction process also indicates that the convergence rate of MPB is slower than the HSPB. But as computer developing, it is not important problem and can be improved by increasing subset number. In this paper, effective hole length is also used in the calculation of the system matrix to correct for the penetration, which has a satisfactory result. From a visual assessment, the reconstructed image may well satisfy clinical criteria and it is therefore worth investigating this novel MPB collimator system for clinical imaging. However, as a new system, MPB collimators have a number of unresolved problems that require further work.

**5. Acknowledgments**

   This work is supported by the Priority Academic Program Development of Jiangsu College Education.